\documentclass[10pt,letterpaper]{article}
\usepackage{opex3}

\usepackage{subfigure}

\begin{document}

\title{Nonlinear properties of split-ring resonators}

\author{Bingnan Wang$^1$, Jiangfeng Zhou$^1$, Thomas Koschny$^{1,2}$ and \\
  Costas M. Soukoulis$^{1,2,*}$}
\address{$^1$
Ames Laboratory and Department of Physics and Astronomy, Iowa State
University, Ames, Iowa 50011, USA\\$^2$
Institute of Electronic Structure and Laser, FORTH, and Department
of Materials Science and Technology, University of Crete, 71110 Heraklion, Crete, Greece
}%
 \email{$^*$soukoulis@ameslab.gov}
\begin{abstract}
In this letter, the properties of  split-ring resonators (SRRs) loaded with high-Q capacitors and nonlinear varactors are theoretically analyzed and experimentally measured. We demonstrate that the resonance frequency $f_m$ of the nonlinear SRRs can be tuned by increasing the incident power.  $f_m$ moves to lower and higher frequencies for the SRR loaded with one varactor and two back-to-back varactors, respectively. For high incident powers, we observe bistable tunable metamaterials and hysteresis effects.  Moreover, the coupling between two nonlinear SRRs is also discussed.
\end{abstract}
 % REPLACE WITH CORRECT OCIS CODES FOR YOUR ARTICLE

%%%%%%%%%%%%%%%%%%%%%%%%%%%%%%%%%%%%%%%%%% References %%%%%%%%%%%%%%%%%%%%%%%%%

%%%%%%%%%%%%%%%%%%%%%%%%%%  body  %%%%%%%%%%%%%%%%%%%%%%%%%%

\section{Introduction}
Electromagnetic Metamaterials are periodically arranged artificial structures that show peculiar behaviors such as negative refraction not seen in natural materials\cite{smith, soukoulis}.  The photonic atoms, or the element structures of the metamaterials, are typically much smaller in size than the working wavelengths such that the metamaterials can be considered to be homogeneous and macroscopic parameters such as electrical permittivity  and magnetic permeability can be used to describe the electromagnetic properties of the metamaterials \cite{pendry1}. By carefully engineering the photonic atoms, both the permittivity and permeability can be made negative such that negative refraction can be achieved from the metamaterial \cite{veselago, pendry2}. The most widely used structures for this purpose is the composite of short metallic wires and split-ring resonators (SRRs) \cite{parker, nr-exp}. While the array of short wires gives a negative permittivity in the wide frequency range below the effective plasma frequency, the array of SRRs gives a negative permeability in the narrow frequency range just above the resonance frequency so that the effective index of refraction can be negative in a narrow frequency band. While most of the research in this area is in a  linear regime, where the electromagnetic responses are independent of the external fields, some effort has  been made to study the nonlinear effects of the metamaterials, especially the nonlinear tunability of the SRRs [8-12]. The SRRs are essentially LC resonators and the resonance frequency is determined by the geometry of the rings. To tune the magnetic responses of the SRRs, extra components or materials need to be introduced into the SRRs.

 In Ref. \cite{ozbay}, ferroelectric films are added to the substrate of the SRRs and the magnetic tunability is achieved by controlling the electric permittivity of the ferroelectric films with the change of temperature. In Ref. \cite{smith-nonlinear}, low-doped semiconductors are photodoped within the slits of the SRRs  and the magnetic response is tuned by varying the conductivity of the semiconductors with an external light source.  In Ref. \cite{zhou-nonlinear}, ferrite rods are introduced to ambient the SRR unit cells and the magnetic resonance is modulated by magnetically tuning the inductance of the ferrite rods by an external magnetic field. Compared to these methods, the use of varactors is more feasible in microwave applications in that the tunability can either be realized by a small DC bias voltage \cite{Kivshar-2, martin, wu-nonlinear} or self-tuned by the intensity of the applied electromagnetic fields without biasing \cite{Kivshar-3,hbv}.  Tunable metamaterials, based on the nonlinear SRRs with varactors, have been tested experimentally in both transmission line form \cite{martin, Kivshar-4} and bulk form \cite{wu-nonlinear}.

In this letter, experiments are completed to analyze the properties of the SRRs loaded with linear and nonlinear elements. The multivalue effect \cite{Kivshar-3,Kivshar-4} of nonlinear SRRs with one varactor is analyzed. The nonlinear properties of SRRs with one varactor and two varactors are compared. The coupling effect of two nonlinear SRRs are also discussed.

\section{Nonlinear properties of a single nonlinear SRR}

The geometry of the SRRs is a single ring with the outer diameter of $7$ mm, the inner diameter of $6$ mm, a slit width of $0.7$ mm, and fabricated on a PC board substrate of $0.8$ mm thick. The resonance frequency of the SRR is around 4 GHz. When a high Q capacitor with capacitance 2 pF is mounted into the slit of the SRR, the resonance frequency is brought down to 0.9 GHz. This makes the ratio between the wavelength and the SRR size as large as 50, which is useful to miniaturize the size of potential microwave devices \cite{ozbay-linear}.

\begin{figure}
\begin{center}
  \includegraphics[width=0.75\textwidth]{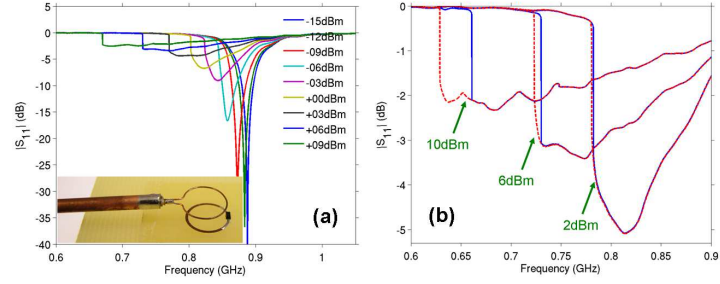}
  \end{center}
  \caption{ (a) The reflection of the SRR loaded with one varactor, measured by a loop antenna. The curves show the measurement results at input power from -15 dBm to 9 dBm in 3 dBm steps. The inset shows the picture of the loop antenna and the sample. (b) The hysteresis effect at high input power levels. The blue curves are measured for forward sweep (the source frequency of the network analyzer is scanned from low to high) and the red dashed curves are measured for reverse sweep (frequency is scanned from high to low).}\label{fig1}
\end{figure}

To study the nonlinear magnetic response of the SRR, a varactor is mounted onto the slit of the SRR. The reflection is measured with a loop antenna and a vector network analyzer. The small loop antenna is made by a semi-rigid coaxial cable and the diameter of the loop is 8 mm. The antenna is kept 2 mm away and  on top of the SRR (see the inset of Fig. \ref{fig1}(a)). The measured reflection at different input power levels is shown in Fig. \ref{fig1}(a). As the incident power from the vector network analyzer increases, the resonance frequency of the nonlinear SRRs moves to lower frequencies. We will able to tune the resonance frequency of the SRR by $ 10 \%$ by increasing the incident power by $6$ dB. If the incident power increases more, one can see clearly from Fig. \ref{fig1}(b) that the reflection coefficient starts showing jumps. In addition, we have observed strong hysteresis and bistable behavior in the nonlinear SRRs. (As one can see from Fig.  \ref{fig1}(a), as the incident power increases the resonance frequency shifts to lower frequencies and, in addition, we observe a decrease in the quality factor of the nonlinear SRR.)

Below, we present some analytical arguments to explain the tunability and the hysteresis effects of the nonlinear SRRs  observed experimentally.

 The varactor,  mounted onto the slit of the SRR, is a Skyworks SMV1231-079 hyperabrupt tuning  varactor.  The nonlinear voltage-dependent depletion layer capacitance $C(V_D)$ ($V_D$ is the voltage across the diode) is described as the following, provided by the manufacturer SPICE model. $ C(V_D)=C_0 \left(1-V_D/V_p\right) ^{-M}$, where $C_0=2.2$ pF is the DC rest capacitance, $V_p=1.5$ V is the intrinsic potential and $M=0.8$. The dissipative current is given by $I_D(V_D)=I_0 \left(e^{(V_D/V_T)}-1 \right)$. From $C(V_D)=\frac{\, dQ_D}{\, dV_D}$, we can determine the time-dependent charge, $ Q_D=\frac{C_0V_p}{1-M}\left[ 1-\left(1-V_D/V_p\right)^{1-M} \right]$. Assume $V_D<V_p$, then the voltage across the diode can be expressed by the charge, $V_D(q)=V_p \left[ 1-\left(1-q \frac{1-M}{V_p}\right) ^{\frac{1}{1-M}} \right]$, where the renormalized voltage is defined as $q=Q_D/C_0$.  

The SRR can be modeled as an RLC circuit with external excitation and the voltage equation can be expressed by $-L \frac{\, dI}{\, dt}-R_S-V_D = \varepsilon(t)$, where $I$ is the current in the resonator,  $L$ is the inductance of the resonator determined by the ring geometry, $R_S$ is the resistance,  and $\varepsilon$ is the driven term provided by the loop antenna in the experiment. For small excitations, $I_D$ can be neglected so  the current can be estimated by $I \approx \, dQ_D/\, dt$. The equation of motion is now $\frac{\, d^2q}{\, dt^2}+ \gamma \frac{\, dq}{\, dt} + \omega_0^2 V_D = -\omega_0^2 \varepsilon(t)$, where $\omega_0^2 = 1/(LC_0)$  and $\gamma = \omega_0^2 R_S C_0$. Expand the restoring term $V_D$ by the Taylor series for small oscillations (the oscillation amplitude satisfies $(1-M)|q|<V_p$) and omit the higher order terms, $V_D(q) \approx q - \frac{M}{2V_p}q^2 + \frac{M(2M-1)}{6V_p^2}q^3$. Assume harmonic excitation so that $\varepsilon (t) = f \cos (\omega t)$, where $f$ is the excitation amplitude and $\omega$ is the excitation frequency,  the equation of motion is further estimated by $\frac{\, d^2q}{\, dt^2}+ \gamma \frac{\, dq}{\, dt} + \omega_0^2 q + \alpha q^2 +\beta q^3 = -\omega_0^2 f \cos (\omega t)$, where $\alpha = - \frac{\omega_0^2 M}{2 V_p}$, $\beta = \frac{\omega_0^2 M (2M-1)}{6V_p^2}$. This is now a nonlinear driven oscillator problem \cite{landau}.

The driven frequency  can be written as $\omega = \omega_0 + \delta$. When $\delta$ is small, the driven frequency is close to the resonance frequency. Without the $q^2$ and $q^3$ term, the oscillator is linear and the amplitude of oscillation, $b$, is given by $b^2 \left( \delta ^2 + \gamma ^2/4 \right) = \omega_0^2 f^2/4$. The nonlinear $q^2$ and $q^3$ terms make the eigen-frequency amplitude dependent, $\omega_0 \to \omega_0 + \kappa b^2$, where $\kappa = \frac{3 \beta}{8 \omega_0} - \frac{5 \alpha ^2}{12 \omega_0^3}$. Then $\delta \to \delta - \kappa b^2$, and the oscillation amplitude satisfies the equation  $b^2 \left[ \left(\delta - \kappa b^2 \right) ^2 + \gamma ^2/4 \right] = \omega_0^2 f^2/4$. This is a cubic equation about $q^2$ and the real roots give the amplitude of oscillations. When the external excitation is small, the oscillation amplitude is also small and the higher orders of $b$ may be neglected and the oscillation can be considered to be linear. When the excitation power is larger, the curve is distorted and the resonance shifts to a lower frequency, since, in our case, $\kappa$ is negative. When the excitation power is large enough, there are three real roots of $b^2$ and the curve is folded over, see Fig. \ref{fig1}(b). The branch in the middle is unstable and the oscillation tends to go to the other two branches. In experiment, the oscillation follows the lower branch until it jumps to the higher branch for forward sweep, and follows the higher branch until it jumps to the lower branch. So, the hysteresis effect is observed in our experimental measurements. Note,  when the voltage on the varactor is larger than 0.5 V, the nonlinear DC dissipative current ( $I_D(V_D)=I_0 \left( e^{V/V_T}-1\right)$, where $V_T=k_B T/e$ is the thermal voltage,  $k_B$ is the Boltzmann constant, $T$ is the temperature, and $e$ is the electron charge) sets in and this increases the loss on the SRR and the reflection dip measured is not as strong as the  small oscillation case. See Fig. \ref{fig1}(a), the reflection minimum increases from around -40 dB to -3 dB when the input power is increased from -15 dBm to 9 dBm. This is not covered in the simplified model of the nonlinear oscillator model.

\begin{figure}
\begin{center}
  \includegraphics[width=0.45\textwidth]{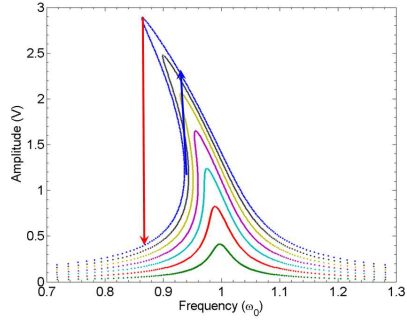}\\
  \end{center}
  \caption{The oscillation amplitude vs. source frequency calculated from the simplified nonlinear oscillator model. The curves from bottom to top correspond to excitation power from low to high. The blue arrow shows the jump for forward sweep of the top curve and the red arrow shows the jump for reverse sweep of the top curve.}\label{foldover}
\end{figure} 

\begin{figure}
\begin{center}
  \includegraphics[width=0.45\textwidth]{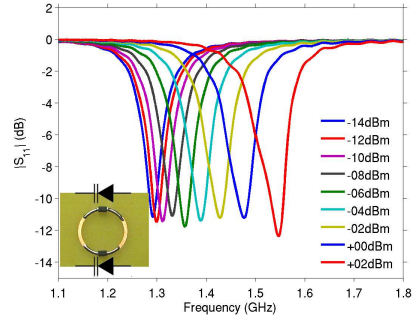}\\
  \end{center}
  \caption{The reflection of the SRR loaded with two back-to-back varactors, measured by a loop antenna. The curves show the measurement results at power levels from -14 dBm to 2 dBm, in 2 dBm steps. The inset shows the sample.}\label{b2b}
\end{figure}

To remove the nonlinear DC current and obtain a  better self-tuning effect, a new SRR is fabricated. The SRR ring is of the same geometry, but with another identical cut on the other side of the ring and a varactor is mounted onto each of the cuts, see inset in Fig. \ref{b2b}. The varactors are arranged back-to-back such that no DC current can circulate in the SRR and the effective capacitance characteristics $C(V)$ of the two varactors is now symmetric. This configuration has the same effect of one heterostructure barrier varator (HBV) diode \cite{hbv}. The two varactors can now be regarded as two tunable capacitors connected in a series. The total capacitance is smaller than a single varactor at the same power level,  so that the resonance frequency shifts to a higher region (see Fig. \ref{b2b}). When the input power increases, the effective capacitance decreases and the resonance frequency increases. Also see from Fig. \ref{b2b} that the resonance strength and the quality factor are almost the same for different input power levels. The only problem of this configuration is that the varactors discharge very slowly,  due to the lack of a  circulating current.

\section{Mutual coupling between two nonlinear SRRs}
\begin{figure}
\begin{center}
  \includegraphics[width=0.7\textwidth]{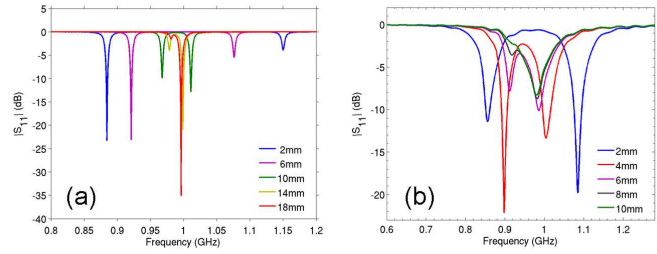}\\
  \end{center}
  \caption{The reflection of (a) two linear SRRs and (b) two nonlinear SRRs at input power of $-15$ dBm.  The legend shows the distance between the two SRRs. The distance between the first SRR and the antenna is fixed.}
\label{linear vs nonlinear}
\end{figure}

To make nonlinear metamaterials out of the nonlinear SRRs,  the mutual coupling of the varactor loaded SRRs must to be studied. Since the mutual coupling of two coplane SRRs are very weak, we studied the case of two parallel SRRs with the same axis (solenoid case). The loop antenna is also parallel to the two SRRs and has the same axis. The distance of the antenna and the first SRR is fixed, and the second SRR is moved away from the first SRR.  Reflection is measured by the loop antenna for different distances between the two SRRs.  In Fig. \ref{linear vs nonlinear}(a), we present the frequency-dependence of the reflection coefficient for two linear SRRs, as the distance between the two SRRs increases. As one can see from  Fig. \ref{linear vs nonlinear}(a), if the distance between the SRRs is large, only one reflection resonance is observed and as the two SRRs move  closer,  the mutual coupling  becomes  stronger and the reflection splits. The closer the two SRRs, the wider the split. For two linear SRRs, the mutual coupling can be calculated analytically with a  simple LC model. 

\begin{figure}
\begin{center}
  \includegraphics[width=0.75\textwidth]{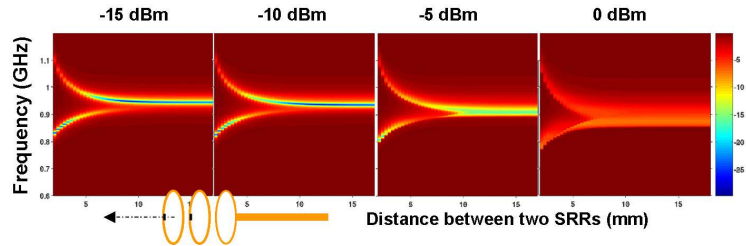}\\
  \end{center}
  \caption{The 2d plot of the reflection of two coupled nonlinear SRRs. The x-axis shows the distance between the two SRRs and the y-axis shows the frequency. The four figures from left to right display the result at input power from -15 dBm to 0 dBm. The SRRs and the antenna are arranged such that they have the same axis and the planes of the rings are parallel to each other. The distance between the loop antenna and the first SRR is fixed and the second SRR is movable along its axis, as seen from the inset. }\label{coupling}
\end{figure}

In Fig. \ref{linear vs nonlinear}(b), we present the frequency-dependence of the reflection coefficient for two nonlinear SRRs, as the distance between the two SRRs increases. The incident power is $-15$ dBm, which is relatively low and the results presented in Fig. \ref{linear vs nonlinear}(b) are almost equivalent to the linear SRRs presented in Fig. \ref{linear vs nonlinear}(a), except that the reflection dip is not as deep as the linear case, due to a higher loss in the varactor than the high-Q capacitor.    When the input power is low, the nonlinear SRRs behaves like the linear ones and the splitting is nicely seen in the -15 dBm case of the 2d plot in Fig. \ref{coupling}. When the input power gets higher, the resonance of the SRR gets broadened, due to higher loss  and the splitting becomes worse, as seen in Fig.  \ref{coupling}. One can clearly see from  Fig.  \ref{coupling} that the hybridization gets very weak as the incident power increases.

\section{Conclusion}
In conclusion, we have demonstrated experimentally dynamic tunability, hysteresis, and bistable behavior in nonlinear SRRs. The nonlinear SRR has a typical SRR design and we have soldered in the gap of the SRR a commercial varactor diode. Tunability of the resonance frequency was completed dynamically by increasing the incident power of the vector network analyzer. We have introduced different nonlinear designs and observed experimentally that the resonance frequency can decrease or increase by increasing the incident power. This way, we are able to change the sign of the nonlinearity. A theoretical model was given that explained all the observed nonlinear effects. Finally, we study the hybridization effects of the linear and nonlinear SRR.

\section*{Acknowledgments}

Work at Ames Laboratory was supported by the Department of Energy (Basic Energy Sciences) under contract No. DE-AC02-07CH11358. This work was partially supported by the office of Naval Research (Award No. N00014-07-1-D359), and European Community FET project PHOME (Contract No. 213390).

%\end{acknowledgments}

\end{document}